\pdfoutput=1
\documentclass[runningheads]{llncs}
\usepackage{algorithm}
\usepackage{algorithmic}
\usepackage{graphicx}
\usepackage{amsmath}
\usepackage{amsfonts,amssymb}
\begin{document}
\title{Understanding Auditory Evoked Brain Signal via Physics-informed Embedding Network with Multi-Task Transformer}
%
\author{Wanli Ma\inst{1} \and
Xuegang Tang\inst{2} \and
Jin Gu\inst{3} \and
Ying Wang\inst{4} \and
Yuling Xia \inst{1}}

\authorrunning{Wanli Ma et al.}
\institute{School of Mathematics, Southwest Jiaotong University, Sichuan, Chengdu, China 
\email{wanlim@my.swjtu.edu.cn}\quad \email{xyl2024001@my.swjtu.edu.cn} \and
School of Computer, SWJTU-Leeds Joint School, Sichuan, Chengdu, China
\email{xuegangt@my.swjtu.edu.cn}\and
School of Computing and Artificial Intelligence, Southwest Jiaotong University, Sichuan, Chengdu, China\\
\email{gujin@swjtu.edu.cn} \and
Chengdu University of Technology Oxford Brookes College, Sichuan, Chengdu, China\\
\email{1428290801@qq.com}}
\maketitle              
\begin{abstract}
In the fields of brain-computer interaction and cognitive neuroscience, effective decoding of auditory signals from task-based functional magnetic resonance imaging (fMRI) is key to understanding how the brain processes complex auditory information. Although existing methods have enhanced decoding capabilities, limitations remain in information utilization and model representation. To overcome these challenges, we propose an innovative multi-task learning model, Physics-informed Embedding Network with Multi-Task Transformer (PEMT-Net), which enhances decoding performance through physics-informed embedding and deep learning techniques. PEMT-Net consists of two principal components: feature augmentation and classification. For feature augmentation, we propose a novel approach by creating neural embedding graphs via node embedding, utilizing random walk to simulate the physical diffusion of neural information. This method captures both local and non-local information overflow and propose a position encoding based on relative physical coordinates. In the classification segment, we propose  adaptive embedding fusion to maximally capture linear and non-linear characteristics. Furthermore, we propose an innovative parameter sharing mechanism to optimize the retention and learning of extracted features. Experiments on a specific dataset demonstrate PEMT-Net’s significant performance in multi-task auditory signal decoding, surpassing existing methods and offering new insights into the brain’s mechanisms for processing complex auditory information.

\end{abstract}
\section{Introduction}
Brain-computer interaction (BCI), which utilizes subtle changes in brain activity and transforms these neural signals into executable computational instructions, shows great potential for deepening our understanding of brain mechanisms. Although this field is still in its infancy, it has already shown promising applications in clinical medicine, biomedicine, and other fields. Current research has focused on visual and spatial cognition\cite{in2} or motor imagery, highlighting the important need for auditory-focused research which has not been done much. Task-state functional magnetic resonance imaging (ts-fMRI), which is characterized by high spatial resolution and non-invasiveness, is pivotal in cognitive neuroscience. Recently, functional connectivity (FC) analysis has become one of the most commonly used methods for describing brain functioning\cite{in4}, and previous studies have shown that functional connectivity is a potential brain marker for predicting cognitive and behavioral traits\cite{in5,in6,in7},
However, the complexity of brain networks poses a great challenge to traditional machine learning models such as support vector machines and logistic regression; moreover, the reliance of these algorithms on manual feature extraction limits their ability to learn intrinsic data features, which in turn affects the model's generalization ability\cite{in10}.

Data augmentation is a key technique to overcome data scarcity and improve model robustness and training efficiency\cite{dataAug1}
to improve model training effectiveness\cite{dataAug}. Fahimi et al. proposed a deep convolutional generative adversarial network-based framework for data augmentation, which achieved better results\cite{DCGANs}. However, traditional data enhancement techniques may introduce artificial variations that do not necessarily match the natural pattern of neural activity\cite{artificial}. In addition, they may not be able to effectively capture the complex linear and nonlinear intricacies of the complex relationships between different brain regions\cite{nonlinear}. Deep learning methods have become cutting-edge approaches for analyzing fMRI datasets, successfully extracting meaningful information from complex connectivity patterns in neuroimaging studies of cognitive abilities \cite{in12,in13}. Gao et al. proposed a group quadratic graph convolutional network that improves the ability of individual neurons to represent complex data \cite{gcn}. However, methods like this do not maximize the capture and utilization of linear and nonlinear features. In addition, the complexity of brain signals and the intricate topology of brain networks pose a great challenge to model generalization\cite{comp}. Models trained in isolated tasks often have difficulty adapting to new unknown conditions due to the inability to cover all neural connections and interaction patterns.

To address the aforementioned challenges, we propose a novel model that innovatively leverages physical information to enhance embedding representations, fostering deep information sharing across tasks and thereby improving the model's generalizability. Our main contributions are as follows:

(1)Proposing PEMT-Net, a groundbreaking multi-task neural network model that leverages physical information embedding and simulates the neural diffusion process with random walks, adeptly capturing extensive neural interactions and introducing a novel method for constructing neural embedding maps from graph features.(2)Our innovative approach includes a unique physical location encoding strategy that delineates the spatial relationships of brain regions based on their correlation, enabling precise physical positioning within the neural network's structure.(3)By integrating adaptive embedding fusion technology and fine-grained multi-task parameter sharing, PEMT-Net achieves a deep understanding of both linear and non-linear features, substantially improving model generalizability across diverse cognitive tasks.

\section{Methods}
\subsection{Overview Methods}
\begin{figure}[htbp]
\centering
\includegraphics[scale=1.05]{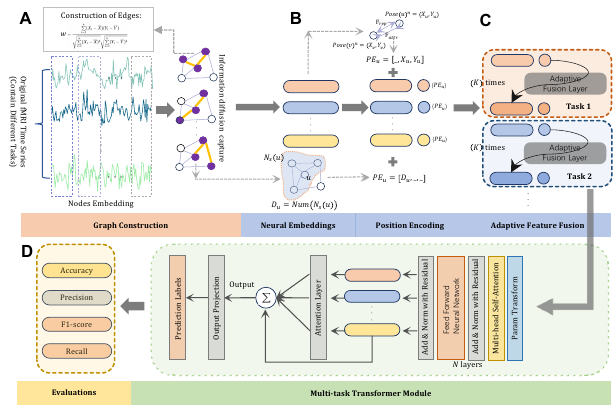}
\caption{Flowchart of our PEMT-Net Model}
\label{fig1}
\end{figure}

In this study, we propose a physics-informed embedding network with multi-task transformer (PEMT-Net)-an innovative model for accurately decoding auditory signals, as shown in Fig. \ref{fig1}, which consists of two main parts:(1) feature augmentation, and (1) classification. In Part 1, neural diffusion process is utilized to distill the neural embedding representation of the region of interest (ROI) from the physical dimension, followed by the fusion of degree centrality and location coding computed by Fruchterman-Reingold algorithm to form rich spatial feature information. In Part 2, adaptive embedding fusion is used to enhance the feature representation, and a soft parameter-sharing mechanism is introduced to support efficient multi-task learning by front-loading the parameter transformation layer in the Transformer encoder. The flowchart of our model is shown in \textbf{Fig. \ref{fig1}} below.

\subsection{Problem definition}

In order to express our actual work in mathematical expression, let $G=\{G^1,G^2,\\\ldots,G^T\}$,  where $G^j=\left(V^j,E^j,W^j\right)$. A subgraph $G^j$ represents a subject's brain region connectivity at a fine granularity. Every Region of interests (ROIs) is considered as a node $u^j\in V^j$, and every edge among them is represented as $e\left(u^j,v^j\right)\subset E^j$, with a weight $w_{u^jv^j}\in W^j$.

\subsection{PEMT-Net Model}
\subsubsection{Neural Diffusion Process.} We apply the core idea of node2vec\cite{node2vec}. As shown in \textbf{Fig.\ref{fig1} A}, the inputs of our first layer are bags of nodes and their weight edges, and the model is trained only on each bag respectively to capture the neural diffusion situation inside the brains. For every node $u^j\in V^j$, let $N_s^j\left(u^j\right)\subset V^j$ be a neighborhood set of node $u^j$, which is formed by applying a neighborhood sampling strategy denoted as $S$.
The goal of this feature learning framework is to learn a mapping function $f^j:V^j\rightarrow \mathbb{R}^d$. This is achieved by maximizing the co-occurrence probability of nodes in a sequence of nodes using the Skip-gram model, and the objective function is defined as:
\begin{equation}
    \max_{f^j}{\sum_{v^j\in V^j} \log{Pr\left(N_s^j\left(u^j\right)\middle| f^j\left(v^j\right)\right)}}
\end{equation}
where the conditional probability is defined by softmax function:
\begin{equation}
    Pr\left(u^j\middle| v^j\right)=\frac{\exp{\left(f^j\left(u^j \right)\cdot f^j\left(v^j\right)\right)}}{\sum_{n\in V^j} \exp{\left(f^j\left(n\right)\cdot f^j\left(v^j\right)\right)}}
\end{equation}

Our algorithm prominently features weighted random walk, leveraging the graph's edge weights to produce a node sequence. This mechanism aims to determine the subsequent node at every step, considering the current node's neighbors and the edge weights to maintain a balance between exploration and exploitation\cite{xia}. This is achieved through the adjustment of parameters $p$ and $q$, which are instrumental in regulating this balance. Assume that the node is located at node $u^j$ at time $t$, the probability of moving to node $v^j$ next is defined as 
\begin{equation}
    P\left(u^j,v^j\right)=\frac{\alpha_{pq}\left(u^j,v^j\right)\cdot w_{u^jv^j}}{\sum_{n\in N_s^j\left(u^j\right)}{\alpha_{pq}\left(u^j,v^j\right)}\cdot w_{u^jn}}
\end{equation}
where $\alpha_{pq}\left(u^j,v^j\right)$ is a preference weight adjusted to $p$ and $q$ that balances the propensity to return to previous nodes with the propensity to explore new nodes, which is defined as
\begin{equation}
\alpha_{pq}(u^j,v^j)=
    \begin{cases}
        \frac{1}{p}&\text{if } v=prev,\\
        1&\text{if } v^j \in U_s(v^j),\\
        \frac{1}{q} &\text{if } v^j \notin U_s(v^j),  v^j \in U_s^{(2)}(v^j).
    \end{cases}
\end{equation}

Finally, the node embedding obtained by the above method is denoted as $X\in \mathbb{R}^{Num(V)\times d}$.
\subsubsection{Physical Position Encoding.} As we describe in \textbf{Fig. \ref{fig1} B}, we define two types of location coding, based on the degree centrality of the nodes and the coordinates of the nodes computed by the FR algorithm \cite{FR}, respectively. Let ${PE}_u = [D_{u},X_{u},Y_{u}]$, where $D_u$ is the degree of node $u\in V$, which is calculated by
\begin{equation}
    D_u = \frac{Num(N_s(u))}{Num(V)-1}
\end{equation}
$X_{u},Y_{u}$ are the coordinates of the FR algorithm. 

In our module, the process of FR algorithm corresponds to weight. We define repulsive force as normal:
\begin{equation}\label{eq1}
    F_{\text{rep}}(u,v)=-\frac{k^2}{||pos(u)-pos(v)||}
\end{equation}
where $pos(u),pos(v)$ denotes the position of node $u,v\in V$, but define attractive force with weight:
\begin{equation}\label{eq2}
    F_{\text{attr}}(u,v) = \frac{||pos(u)-pos(v)||^2}{k}\cdot w_{uv}
\end{equation}
which means the higher the weight, the stronger the attraction, the nodes connected by that edge will be pulled closer together. The algorithm is demonstrated below:
\begin{algorithm}
	\renewcommand{\algorithmicrequire}{\textbf{Input:}}
	\renewcommand{\algorithmicensure}{\textbf{Output:}}
	\caption{Fruchterman-Reingold Algorithm}
	\label{alg1}
	\begin{algorithmic}[1]
            \REQUIRE weight matrix $\{W_{uv}\}^j$
		\STATE Initialization:$\{pos(v)\}$ for all $v\in V$, $k=\displaystyle{\sqrt{\frac{area}{|V|}}}$, $maxIterations$, $n\leftarrow 1$
		\REPEAT
		\STATE $n \leftarrow n + 1$
		\STATE Repulsion Step: for each pair of nodes $(u,v),u\neq v$: \\$\displaystyle{\Delta pos(u)^{n+1}+=\frac{pos(u)^n-pos(v)^n}{||pos(u)^n-pos(v)^n||}\times F_{\text{rep}}(u,v)}$ 
		\STATE Attraction Step: for each edge $(u,v)\in E:$\\$\displaystyle{\Delta pos(u)^{n+1} -= \frac{pos(u)^n-pos(v)^n}{||pos(u)^n-pos(v)^n||}\times F_{\text{attr}}(u,v)}$\\
  $\Delta pos(v)^{n+1}+=$
  $\displaystyle{\frac{pos(v)^n-pos(u)^n}{||pos(u)^n-pos(v)^n||}\times F_{\text{attr}}(u,v)}$
		\STATE Position Update: for each node $v\in V$:\\
  $\displaystyle{pos(v)^{n+1}+=\frac{\Delta pos(v)^{n+1}}{||\Delta pos(v)^{n+1}}\times \min{(||\Delta pos(v)^{n}||,t)}}$
            \STATE Cooling: $t-=\Delta t$
		\UNTIL $t = maxIterations$
		\ENSURE  optimized graph layout $\{pos(v)\}$ for all $v\in V$
	\end{algorithmic}  
\end{algorithm}

In the end of this part, the original embedding is directly aggregated with the physical location encoding, which is denoted as:
\begin{equation}
   X'= X || \{{PE}_u\}_{u\in V}
\end{equation}
where notation $||$ is the symbol of concatenation.
\subsubsection{Adaptive Embedding Fusion.} In this section, we propose a multi-round adaptive embedding fusion method to enhance feature representation, which is shown in \textbf{Fig. \ref{fig1} C}. Consider the graph given above $G = \{G^j\}_{j\in T}$, where weight matrix $W\in M_N$, and feature matrix $X'\in M_{N,{d+3}}$. In the case of $k$ rounds of the feature propagation process, the feature update of a node can be expressed as:
\begin{equation}
    X'^{(k+1)} = W\times(X'^{(k)}\odot W')+X'^{(k)}
\end{equation}
where $X'^{(k+1)}$ is the feature matrix after the $k$-th round of update, $\odot$ denotes element-by-element multiplication, and $W'$ is the weight matrix calculated based on the node degree for realizing adaptive embedding fusion.

The feature after $k$-th round propagation enhancement is $S \in M_{N,k+1,d}$.

\subsubsection{Multi-Task Transformer.} In this section, we propose a transformer model that can deal with multi-task missions, which is demonstrated in \textbf{Fig. \ref{fig1} D}. The core component of the transformer is self-attention mechanism \cite{trans}. As our input matrix is $S$, then the output matrix of single-head attention is calculated as:
\begin{equation}
    Attention(Q,K,V) = \text{softmax}\left(\frac{QK^\mathrm{T}}{\sqrt{d_K}}\right)V
\end{equation}
where $Q = SH^Q, K = SH^K, V = SH^V$, and $H^Q\in \mathbb{R}^{d\times d_Q}, H^K\in \mathbb{R}^{d\times d_K},H^V\in \mathbb{R}^{d\times d_V}$ are projection matrices. And for multi-head attention, similarly,
\begin{equation}
    MultiHead(Q,K,V) = Caoncat(head_1,\dots,head_n)H^O
\end{equation}
where
\begin{equation}
    head_i = Attention(QH_i^Q,KH_i^K,VH_i^V)
\end{equation}
The self-attention mechanism in each encoder layer is followed by a feed-forward neural network that further processes the output of the self-attention layer. This network usually contains two linear transformations and a nonlinear activation function of the following form:
\begin{equation}
    \text{FFN(x)} = \max{(0,xW_1+b_1)W_2+b_2}
\end{equation}
where $W_1,W_2$ are weight matrix, $b_1,b_2$ are terms of bias.

In addition, in order to adapt to multi-task learning and share parameters, we include a parameter transformation layer in front of each encoder layer, which allows different parts of the model to share the same base parameters in a transformed form for soft parameter sharing. Let the parameters used in the encoder be $\Theta$. The parameter transformation layer in front of each encoder layer can be represented by a linear transformation that corresponds to a weight matrix $W_i$ and a bias vector $b_i$, where $i$ denotes the index of the encoder layer. Thus, the parameter transformation can be represented as:
\begin{equation}
    \Theta_i'=W_i\Theta+b_i
\end{equation}
where $\Theta_i'$ is the transformed parameter, which will be used in $i$-th encoder layer. In this way, although the parameters used in the encoder layer are transformed, the basis of these parameters is the same.

Furthermore, inside each encoder layer, $Q, K$ and $V$ in the self-attention mechanism can be calculated using the transformed parameters: 
    \begin{align}
        Q_i = S_iW_i^Q\Theta_i'\\
        K_i = S_iW_i^K\Theta_i'\\
        V_i = S_iW_i^V\Theta_i'
    \end{align}

To process the data processed by the decoding layer, we apply the attention-based readout layer involved in the model NAGphormer \cite{NAG}, which is represented in \textbf{Fig. \ref{fig1} C}. Up to this point, the physics-informed embedding information extracted from the brain regions has been fully learned by our model.


\section{Experiments}
\subsection{Data and Pre-processing}
The proposed model was evaluated on the fMRI dataset we collected from healthy subjects. In the fMRI experiment,  subjects were asked to imagine and listen to four categories auditory information (Human, Animal, Machine, Nature) , and we obtained eight categories of auditory neural activity which would be classified to test the performance of our model. For fMRI data pre-processing, the first five volumes of each run were discarded \cite{dp}. All images were realigned to remove movement artifact, then coregistered and standardized to MNI space with a voxel size resampled to 3×3×3 mm using the T1 images. The normalized images were smoothed with a 6-mm Gaussian kernel. Then we extract time series to form functional connection.

\subsection{Implementation Details}
During the initial phase of exploration through random walks, we configure the step size of each walk to be 10, covering a total distance of 100. This process yields embeddings with a dimensionality of 256. In the subsequent phase, the model operates with a batch size of 32 and undergoes 7 iterations of adaptive feature fusion. The Transformer architecture incorporates 8 attention heads, and its hidden layer is set to a dimensionality of 512. Moreover, the model features 2 layers dedicated to encoding tasks (with parameter transformation layers). Training is conducted over 200 epochs, with the data split as follows: 70\% for training, 15\% for validation, and the remaining 15\% for testing purposes. The entire model training process is executed on a single M1 GPU. We used accuracy (Pre), F1 score (F1), and Recall score (Recall) to evaluate the model and plotted the confusion matrix to visualize the model classification more intuitively.
\subsection{Results of Fine-Grained Classification}
\begin{figure}[htbp]
\centering
\includegraphics[scale=0.5]{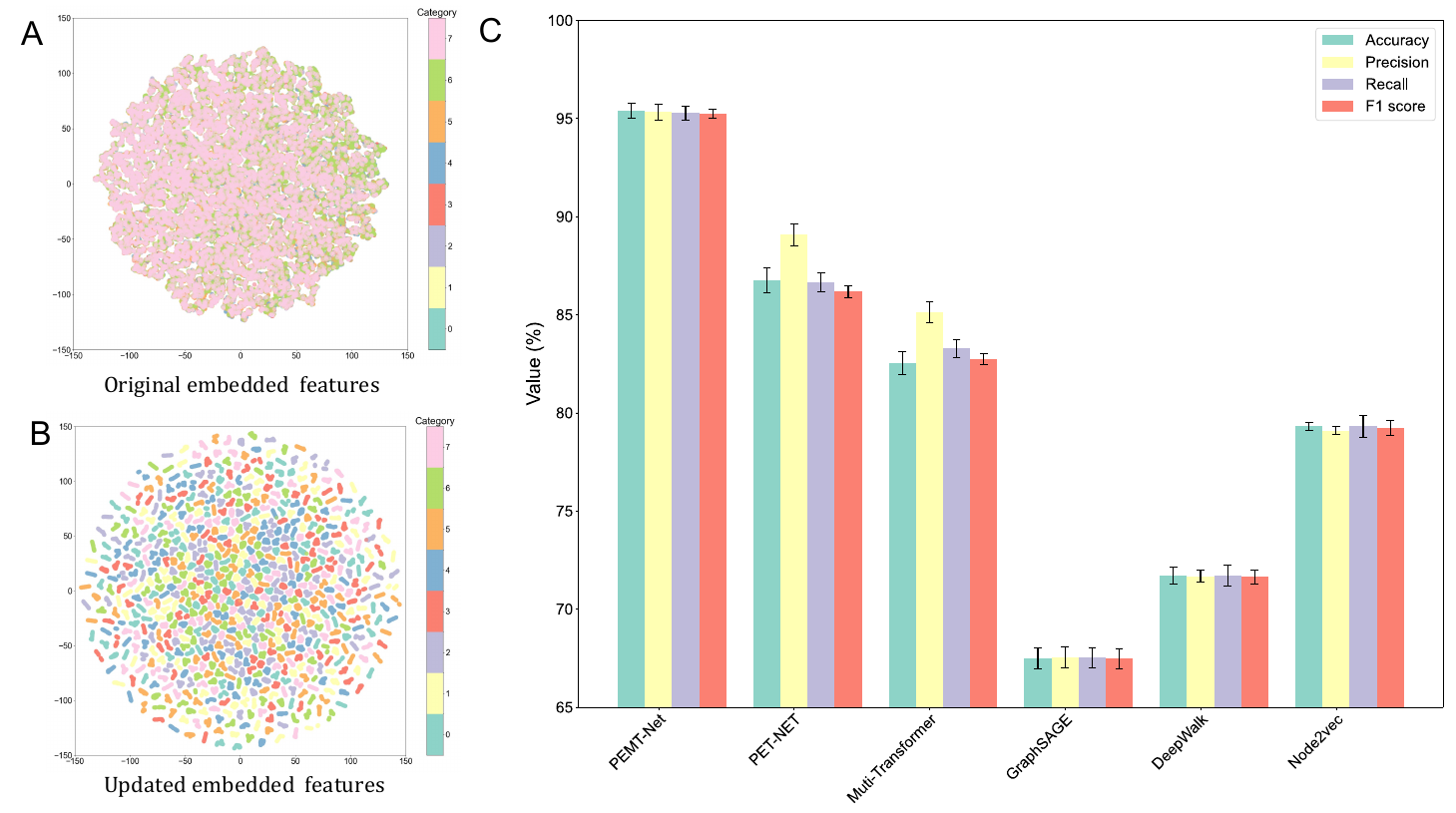}
\caption{\textbf{A} represents the distribution of the original features after dimensionality reduction by the t-SNE method; \textbf{B} represents the distribution of the high-dimensional embeddings encoded with physical locations after dimensionality reduction by the t-SNE method; \textbf{C} represents the error bar graphs of each metric for each method.}
\label{fig2}
\end{figure}

According to \textbf{Fig. \ref{fig2} A and B}, we can observe that the original features exhibit a mixed distribution, with nodes of different classification results chaotically clustered together. In contrast, nodes constrained by physical space and those with captured information overflow exhibit a regular distribution, generally showing clustering of nodes within the same category and almost no clustering between nodes of different categories. Therefore, our model incorporates node embedding of brain neural information, simulating the physical process of information diffusion, capable of capturing both local and non-local overflow effects, thereby benefiting the subsequent classification prediction part of the model.

In this experiment, we performed a fine-grained categorization (8 categories) to validate our model, where there were a total of 4 tasks that were evenly distributed in each category. To benchmark our method, we introduce several baselines: GraphSAGE, DeepWalk, Node2vec, and to verify the feasibility of our idea, we did the following ablation experiments respectively: multi-transformer (without physics-informed), PET-Net (without sharing parameters) and our proposed model.

\begin{table}[h]
\centering
\caption{Comparation of different methods}\label{tab1}
\setlength{\tabcolsep}{1.5mm}{
\begin{tabular}{ccccc}
\hline
Methods                  & Accuracy & Precision       & Recall          & F1 Score        \\ \hline
GraphSAGE                      &  67.48 ± 0.53        & 67.56 ± 0.54                &  67.52 ± 0.53               & 67.47 ± 0.53                \\
DeepWalk                      & 71.69 ± 0.43        &  71.68 ± 0.30               & 71.70 ± 0.53                & 71.63 ± 0.37                \\
Node2vec                 & 79.31 ± 0.22         & 79.10 ± 0.19               & 79.31 ± 0.55                & 79.23 ± 0.38               \\ \hline
\textbf{Multi-Transformer} &  82.52  ± 0.59      & 85.12 ± 0.52                & 83.28 ± 0.45                &  82.74 ± 0.27               \\
\textbf{PET-Net}         & 86.74 ± 0.64        & 89.07 ± 0.55        & 86.64 ± 0.49        & 86.17 ± 0.29         \\
\textbf{PEMT-Net}        & \textbf{95.41 ± 0.38}         & \textbf{95.32 ± 0.42} & \textbf{95.28 ± 0.36} & \textbf{95.26 ± 0.23} \\ \hline
\end{tabular}}
\end{table}

\textbf{Table \ref{tab1}} and \textbf{Fig. \ref{fig2} C} show all the results of the experiments in this paper, where the upper half is the comparison experiment with baseline and the lower half is the ablation experiment. It is easy to see that our proposed method achieves the best
. For baseline, the accuracy improvement is 16.1\% - 27.93\%. From the ablation experiments, we can see that compared to the Multi-Transformer model, our accuracy is improved by 13.51\%, which indicates that our physically enhanced embedding representation has excellent results; compared to the PET-Net model, our accuracy is ahead by 9.09\%, which indicates that the deep learning module involved successfully captures deep linear and nonlinear features and preserves them in multi-task learning, improving the generalization of the model.

\section{Conclusion}
In this paper, we present a novel model for learning auditory signals in the brain. The model innovatively combines physical modeling and deep learning. Experiments demonstrate that the node representation can be significantly enhanced by modeling the diffusion process of neural signals in the brain and obtaining the physical location encoding based on the interrelationships, and the decoding performance can be improved by a parameter transformation layer that can adequately learn the inter-task activities and share the parameter representation. In the future, we will also simulate the diffusion process by better random wandering, and further optimize the physical location encoding to improve the generalization of the model so that it can be applied in other fields.
\section{Acknowledgement}
This work is supported in part by the  Foundation of XXXX.
\bibliographystyle{splncs04}
\bibliography{PEMT-Net/PEMT-Net_Paper}

\end{document}